\documentclass{article}
\usepackage{amssymb}

\usepackage{graphicx}
\usepackage{amsmath}

\input{tcilatex}

\begin{document}

\title{A numerical analysis for G(t) in the Scalar-tensor theory.}
\author{Luis Augusto Trevisan \\
%EndAName
Demat-Universidade Estadual de Ponta Grossa.\\
Ponta Grossa-PR-Brazil- CEP 84030-000\\
latrevis@uepg.br}
\maketitle

\begin{abstract}
We studied the variation of Newton%
%TCIMACRO{\UNICODE{0xb4}}%
%BeginExpansion
\'{}%
%EndExpansion
s gravitational constant $G,$ using the experimental data and the
scalar-tensor theory for gravitation. We search for a fine tunning to match
the datas and find possible scenarios for the history of this constant.
\end{abstract}

\section{Introduction}

Since the thirties, many cosmologists have been suggested \ theories that \
predict the variation of Newton's gravitational ''constant''. \ Among the
more famous are Dirac's Large Number Hypothesis \cite{Dirac}, the
Brans-Dicke theory \cite{BD} and more recently, some model of scalar-tensor
theories (see the work by Barrow \cite{STT}. for a review). The last one can
be seen as a generalization of Brans-Dicke theory.

On the experimental hand, some measurements are performed to study this
variation; for instance, the spacecraft radar ranging \cite{srr} and lunar
laser ranging \cite{llr} . Both experiments give an upper limit for the
ratio $\left| \dot{G}/G\right| $ $\ $that is:

\begin{equation*}
\left| \frac{\dot{G}}{G}\right| <10^{-12}yr^{-1}
\end{equation*}

On despite of \ so especulative ideas, I have seen few numerical analisys
about the evolution of the value Newton's gravitational constant. \ A work
by Bronikov, Melnikov and Novello \cite{BMN} is in this sense. They describe
the upper limit to $\left| \dot{G}/G\right| .$ Barrow and Parsons \cite{BP}
studied the behavior of cosmological models with varying $G$ .They describe
some analytical solutions to the field equations, using the conformal time
coordinates and showed the dependence of the coupling constant $\omega $
with $\phi /\phi _{0}$ in some particular cases of the models.

In the present work, we made a numerical analisys, based in a Scalar-Tensor
theory, with variable $\omega ,$studied by Barrow and others \cite{STT}.
These numerical analisys, togheter with experimental data, could indicate
the more correct theory, the initial conditions of the Universe and the
evolution.The paper is divided in the following sequence: in next section,
we review the model and isolated some variables that will be important in
the sequence. In section 3, we show the numerical methods and hypotheses. In
section 4 we show some graphics, the main point of this work. The last
section is devoted to discussion and concluion.

\noindent

\section{Relativistic scalar-tensor theories}

Newtonian gravitation permits us to ' write in' an explicit time variation
of G without the need to satisfy any further constraint. However, in general
relativity the geometrical structure of space-time is determined by the
sources of mass-energy it contains and so there are further constraints to
be satisfied. Suppose that we take Einstein's equation in the form\ (c$%
\equiv 1$)

\begin{equation*}
\tilde{G}_{b}^{a}=8\pi GT_{b}^{a}
\end{equation*}
where $\tilde{G}_{b}^{a}$ and $T_{b}^{a}$ are the Einstein and
energy-momentum tensors, as usual, but imagine that $G=G(t).$ \ If we take a
covariant divergence $_{;a}$ of this equation the left-hand \ vanishes
because of the Bianchi identities, $T_{b;a}^{a}=0$ if energy-momentum
conservation is assumed to hold, hence $\partial G/\partial x^{a}=0$ always.
In order to introduce a time- variation \ of $G$ \ we need\ to derive \ the
space and time variations from some scalar \ field, $\psi $ which
contributes a stress tensor $\tilde{T}_{b}^{a}(\psi )$ to the right=hand
side of the gravitational field equations.

We can express this structure by a choice of lagrangian, linear in the
curvatura scalar $R$, that generaizes the Einstein-Hilbert lagrangian of
general relativity with,

\begin{equation}
L=-f(\psi )R+\frac{1}{2}\partial _{a}\psi \partial ^{a}\psi +16\pi L_{m}
\label{F1}
\end{equation}
where $L_{m}$ is the lagrangian of matter fields. The choice of function $%
f(\psi )$ defines the theory. When $\psi $ is constant this reduces, after
rescaling \ of coordinates, to the Einstein-Hilbert lagrangian of general
relativity. The formulation introduced by Brans-Dicke \cite{BD} can be
obtained from \ref{F1} by a non-linear tranformation of $\psi $ and $f.$
Define a new scalar field $\phi $ , and a new coupling constant $\omega
(\phi )$ by

\begin{equation}
\begin{array}{c}
\phi =f(\psi ) \\ 
\omega =f/(2f%
%TCIMACRO{\UNICODE[m]{0xb4}}%
%BeginExpansion
{\acute{}}%
%EndExpansion
%TCIMACRO{\UNICODE[m]{0xb4}}%
%BeginExpansion
{\acute{}}%
%EndExpansion
^{2})
\end{array}
\label{f2}
\end{equation}
then \ref{F1} becames

\begin{equation}
L=-\phi R+\frac{\omega (\phi )}{\phi }\partial _{a}\phi \partial ^{a}\phi
+16\pi L_{m}.  \label{F3}
\end{equation}
The Brans-Dicke theory arises as the special case

\begin{equation}
Brans-Dicke:\quad \omega (\phi )=\text{constant};\text{ }f(\psi )\propto
\psi ^{2}.  \label{f4}
\end{equation}

The field equations that arise by varying the action associated with $L$ in 
\ref{F3} with respect to the metric, $g_{ab\text{ }}$and $\phi $ separately,
gives the field equations,

\begin{equation}
\tilde{G}_{ab}=-\frac{8\pi }{\phi }T_{ab}-\frac{\omega (\phi )}{\phi ^{2}}%
\left[ \phi _{,a}\phi _{,b}-\frac{1}{2}g_{ab}\phi _{,i}\phi _{,i}\right]
-\phi ^{-1}\left[ \phi _{,a;b}-g_{ab}\square \phi \right]  \label{f5}
\end{equation}
\begin{equation}
\left[ 3+2\omega (\phi )\right] \square \phi =8\pi T_{a}^{a}-\omega ^{\prime
}(\phi )\phi _{,i}\phi ^{^{\prime }i}  \label{f6}
\end{equation}
where \ the energy-momentum tensor of the matter sources obeys the
conservation equation

\begin{equation}
T_{;a}^{ab}=0.  \label{f7}
\end{equation}
These field equations reduce to those of general relativity when $\phi $
(and hence $\omega (\phi )$ ) is constant, in which case the Newtonian
gravitational constant is defined by $G=\phi ^{-1}$.

\subsection{The Basic hypothesis of our analisys}

In this work work, we will not suppose any trial function $\omega (\phi ),$
hence we will not propose any analitycal solution to the field equation;
instead we calculated $G(t)$, numerically, based in initial values of $\phi ,%
\dot{\phi}$ and of course $\omega (\phi )$ (given by the field equations).
We suppose that the scalar tensor theory can describe the present Universe
quite well, so we have to consider the recent experimental \ data about it.
We also assume that these data are hold, at least, during the last one
billion of years.

Clearly the equation of state of our work must be:

\begin{equation}
p=0,  \label{f8}
\end{equation}
that is a basic fact in the cosmology.

Now is official, the Universe is flat. With this headlines, the Physicsweb
site call the attention to work by Bernardis et al., in the Nature \cite
{Nature} , about the measures of the density of the Universe. This give us
another information to work, the tricurvature:

\begin{equation}
k=0  \label{f9}
\end{equation}

Last, but not least, is the observation that the Universe is expanding \cite
{accelerating}, not only expanding, but also accelerating, that is, the
deceleration parameter $q$ is:

\begin{equation}
q=-\frac{\ddot{a}a}{\dot{a}^{2}}=-1  \label{f10}
\end{equation}
where $a$ is the scale factor, that implies :

\begin{equation}
\frac{\dot{a}}{a}=H  \label{f11}
\end{equation}
where $H$ is the Huble parameter, and hence

\begin{equation}
a=e^{Ht}.  \label{f12}
\end{equation}

The field equations we have to work now are as follow:

For the Robertson-Walker's metric,

\begin{equation}
ds^{2}=dt^{2}-a(t)^{2}\left[ dx^{2}+dy^{2}+dz^{2}\right]  \label{f13}
\end{equation}
we find, then $(L_{m}=0):$

\begin{equation}
H^{2}=\frac{8\pi \rho }{3\phi }-\frac{\dot{\phi}}{\phi }H+\frac{\omega }{6}%
\frac{\dot{\phi}^{2}}{\phi ^{2}}  \label{f14}
\end{equation}

\begin{equation}
\ddot{\phi}+3H\dot{\phi}=-\frac{\dot{\phi}\dot{\omega}}{3+2\omega }+\frac{%
8\pi }{3+2\omega }\left( \rho -3p\right)  \label{F15}
\end{equation}

\begin{equation}
\rho ^{\cdot }+3\frac{\dot{a}}{a}(\rho +p)=0.  \label{f16}
\end{equation}

One can isolate $\omega .$

\begin{equation}
\omega =6\frac{\phi ^{2}}{\dot{\phi}^{2}}\left[ H^{2}-\frac{8\pi \rho
_{o}e^{-3Ht}}{3\phi }+\frac{\dot{\phi}}{\phi }H\right]  \label{f17}
\end{equation}
and $\dot{\omega},$

\begin{equation*}
\dot{\omega}=12\left\{ \frac{\phi }{\dot{\phi}}-\frac{\phi ^{2}\ddot{\phi}}{%
\dot{\phi}^{3}}\right\} \left\{ H^{2}-\frac{8\pi \rho _{0}e^{-3Ht}}{3\phi }+%
\frac{\dot{\phi}}{\phi }H\right\} +
\end{equation*}
\begin{equation}
\frac{2}{3}\frac{1}{\dot{\phi}^{2}}+\left\{ 24\pi \rho _{0}e^{-3Ht}\left[
3H\phi +\dot{\phi}\right] +H\left[ \ddot{\phi}\phi -\dot{\phi}^{2}\right]
\right\}
\end{equation}

Here $\rho _{0}$ is an initial value for the density. \ Putting $\omega $
and $\dot{\omega}$ in \ref{F15}, we can isolate $\ddot{\phi}.$ This is an
important point of this work, we use the diference finite method to
calculated the field variations. \ We can check that:

\begin{equation}
\ddot{\phi}=\frac{2H+8\pi \rho _{0}e^{-3Ht}-7H\dot{\phi}-12(\phi ^{2}/\dot{%
\phi})\left[ H^{2}-8\pi \rho _{0}e^{=3Ht}/(3\phi )+\phi H/\dot{\phi}\right] 
}{1+2H\phi /\dot{\phi}}  \label{F19}
\end{equation}
and can use :

\begin{equation}
\dot{\phi}(t+\Delta t)=\dot{\phi}(t)+\ddot{\phi}_{t}\Delta t  \label{f20}
\end{equation}
In the same way:

\begin{equation}
\phi (t+\Delta t)=\phi (t)+\dot{\phi}\Delta t  \label{f21}
\end{equation}
and we have recorrence. We still need some information, the intial values
for $\phi $ and $\dot{\phi}$ and will discuss it in the next section.

\section{The initial values problem.}

As we can see in \ref{F19}, $\ddot{\phi}$ depends on the density $\rho _{0}$%
, the Hubble parameter $H$, that are constants. $\ddot{\phi}$ depends also
of $\phi $ and $\ \dot{\phi}$ and the choice of these two intial values
determines the evolution of Newton gravitational constant. We remember that:

\begin{equation}
G=\frac{2\omega +4}{2\omega +3}\frac{1}{\phi }  \label{f22}
\end{equation}
Hence $\omega (\phi (t))$ and $\phi (t)$ must result, when we arrive at the
present Universe:

\begin{equation}
G=6.67\times 10^{-8}\mathrm{dyn.cm}^{2}\mathrm{g}^{-2}  \label{F23}
\end{equation}

Note that we start in the past and go to the present time, so we can have
values of $G$ in the past.\ref{F23} is a final value condition. The theory
has two free parmeters, the initial values for $\phi $ and $\dot{\phi}$, and
could have also two constraint, $G$ and $\dot{G},$ but there are few
measures of $\ \dot{G},$ so still there are some freedom in the choice of
initial values. At the end of this work, we show a result that consider the
limit value for $\dot{G}$. In the Brans-Dicke \cite{BD} theory, we have

\begin{equation}
\phi \propto \frac{1}{G}.  \label{f24}
\end{equation}
We use this fact as a guide to the choice of $\phi .$ For $\ \dot{\phi}$ we
use the relation:

\begin{equation}
\dot{\phi}\sim 10^{n}\phi .  \label{f25}
\end{equation}
$n$ is an integer. If we write:

\begin{equation}
\phi =\frac{b}{G}  \label{fphicg}
\end{equation}
where $b$ is a constant, each $n$ value will determine the value of $b$, (or
we can also work in the opposite sense) in order to satisfy relation \ref
{F23}. \emph{This is the important point of this work, we have to choice the
initial numerical values for }$\phi $ \emph{and for }$\dot{\phi}.$ \emph{%
These values held in the past, one billion years ago. So the ratios }$\dot{%
\phi}/\phi $ \emph{discussed here must be considered in the past. }

Finally, for $H$ we use the same value of \cite{BMN} \ that is :

\begin{equation}
H=0.7\times 10^{-10}yr^{-1}  \label{FH}
\end{equation}
$\rho _{0}$ comes from \cite{Wein} and it is :

\begin{equation}
\rho _{0}=2\times 10^{-29}\mathrm{g/cm}^{3}  \label{rho}
\end{equation}
Of course, when considering the past, we have to correct this density.

With these considerations, we have performed some possible situations, that
are presented in the next section.

\section{Results.}

We studied some possible cases, using the datas and methods explained above,
taking care that each step in our program means 10 millions of years. So in
the figures below, 100 means one billion of years ago.

The first case is the following: we have positive values for $\dot{\phi}$,
in a way that:

\begin{equation}
\dot{\phi}=10^{-n}\phi .
\end{equation}

\bigskip We performed our analisys for the case $n=2,3,4,6,8.$ In the figure
1 we present our results for the \ \ \ case $n=1,2,3.$ \ In our picture, the
present time is on 0.0 and the past goes to right.

In figure 2, we studied another values. We use $n=4,6$ and 8 and $\dot{\phi}$
is positive. In the figure 2 we can these results.

As we can see, it%
%TCIMACRO{\UNICODE{0xb4}}%
%BeginExpansion
\'{}%
%EndExpansion
s interesting to work the case n=3. \ So we studied the cases: $n=2.5,$ and
the more interesting:

\begin{equation}
\dot{\phi}=2\times 10^{-3}\phi .
\end{equation}
In this case, the variation of $G(t)$ is slow and the calculated value for $%
\omega (\phi )$ is bigger than 500. This is in agreement with observacional
data \cite{Will}. More exactly, $\omega (\phi )\approx 889$ (for the present
time).\ This means that this is a good starting point to search the fine
tuning of the theory with the experimental \ data. In the figure 3 we have
the results.

For the sake of completeness, we also show the cases where we have a
negative $\dot{\phi}.$

\subsection{\protect\bigskip The fine tuning}

As a final result, after a search for the parameters that hold theory and
exprimental datas, we find that using:

\begin{equation*}
\phi \approx 0.92755/G
\end{equation*}
as the intial value (one billion years ago), and (also for one billion years
ago)

\begin{equation*}
\dot{\phi}=1.754236908125\times 10^{-3}\phi
\end{equation*}
we obtain the one posible value for $G,$ one billion of years ago:

\begin{equation*}
G_{past}\approx 7.54\times 10^{-8}\mathrm{dyn.cm}^{2}\mathrm{g}^{-2}
\end{equation*}
and the value for $\omega ,$ in the past, was, according to our computation

\begin{equation*}
\omega _{past}\approx 8.87
\end{equation*}
and the present value for $\omega $ is, according to our program: 
\begin{equation*}
\omega \approx 8.24\times 10^{21}
\end{equation*}
and finally, the present ratio: 
\begin{equation*}
\frac{\dot{G}}{G}\approx -5.77\times 10^{-13}/yr
\end{equation*}
Note that, because we have so small experimental value for this ratio, we
have to use several digits in $\dot{\phi}$ .

These are the initial results of our proposal of perform an numerical
analisys of the consequences of the scalar tensor theories of gravitation.
We hope that these informations can be used in another investigations of the
cosmological parameters.

\section{Discussion and Conclusion.}

In this work, we consider the scalar-tensor theory of gravitation, \ which
the has the proposal of study of the Neton%
%TCIMACRO{\UNICODE{0xb4}}%
%BeginExpansion
\'{}%
%EndExpansion
s gravitational constant. We didn%
%TCIMACRO{\UNICODE{0xb4}}%
%BeginExpansion
\'{}%
%EndExpansion
t sugest any analitycal solution to the field equations. What we have done
is, based in some justified hypothesis, to calculated numerically the
solutions for the fields ($\phi )$, the coupling constant ($\omega )$, and
the main point, the values of $G(t).$ On despite that the field equations
are non-linear in $\phi $ and $\dot{\phi}$ we have two constraints ($G$ and $%
\left| \dot{G}/G\right| $ )\ that filter the possibilities. It%
%TCIMACRO{\UNICODE{0xb4}}%
%BeginExpansion
\'{}%
%EndExpansion
s useful to study if we can aply the uniquiness of solution under these
conditions.

We consider that numerical analisys presented is quite interesting, because
shows the possible evolutions for the value of the Newton%
%TCIMACRO{\UNICODE{0xb4}}%
%BeginExpansion
\'{}%
%EndExpansion
s gravitational constant $G$ in a recent period of the Universe, based in a
Scalar Tensor Theory of gravitation. The same idea can be used to make some
predictions about future values of G. We believe that, if one consider our
basic hypothesis valid, ($k=0,$ accelerating Universe, $p=0)$ it%
%TCIMACRO{\UNICODE{0xb4}}%
%BeginExpansion
\'{}%
%EndExpansion
s worthwhile to work to find the intial values of $\phi $ and $\dot{\phi}$
that results in the present values of $G$ and $\dot{G}$ (if the last one is
measured). The numerical approach used here can be improved to work with a
more accurate step in the program (less than 10 millions of years). Another
point is that our computational program shows that is possible a change in
the signal of $\dot{G},$ so we may especulate that $G$ can start to grow.

We can also consider that, if the Newton%
%TCIMACRO{\UNICODE{0xb4}}%
%BeginExpansion
\'{}%
%EndExpansion
s gravitational constant have changed in the last one billion of years,
there are consequences and evidences that can be search in the geological or
astronomical fields of knowledge, for a compreensive discussion about these
implications, see \cite{Wein}. Eventually, the study of gravitational waves
can give some information about the value of $G$ in the past, or , in the
opposite sense, these study can use some information about the numerical
evolution of $G,$ gravitational waves in the scalar-tensor theories are
studied in \cite{GWSTT}.

As a final consideration, we remember that we use $L_{m}=0.$ It%
%TCIMACRO{\UNICODE{0xb4}}%
%BeginExpansion
\'{}%
%EndExpansion
s possible propose many different forms for $L_{m}.$ In our humble opinion,
if we think that even the theorethical part is not solved yet, hence the
numerical one, in a way that it%
%TCIMACRO{\UNICODE{0xb4}}%
%BeginExpansion
\'{}%
%EndExpansion
s possible to perform several \ approachs to the problem.

\pagebreak 

\begin{center}
{\Large Captions}
\end{center}

\bigskip 

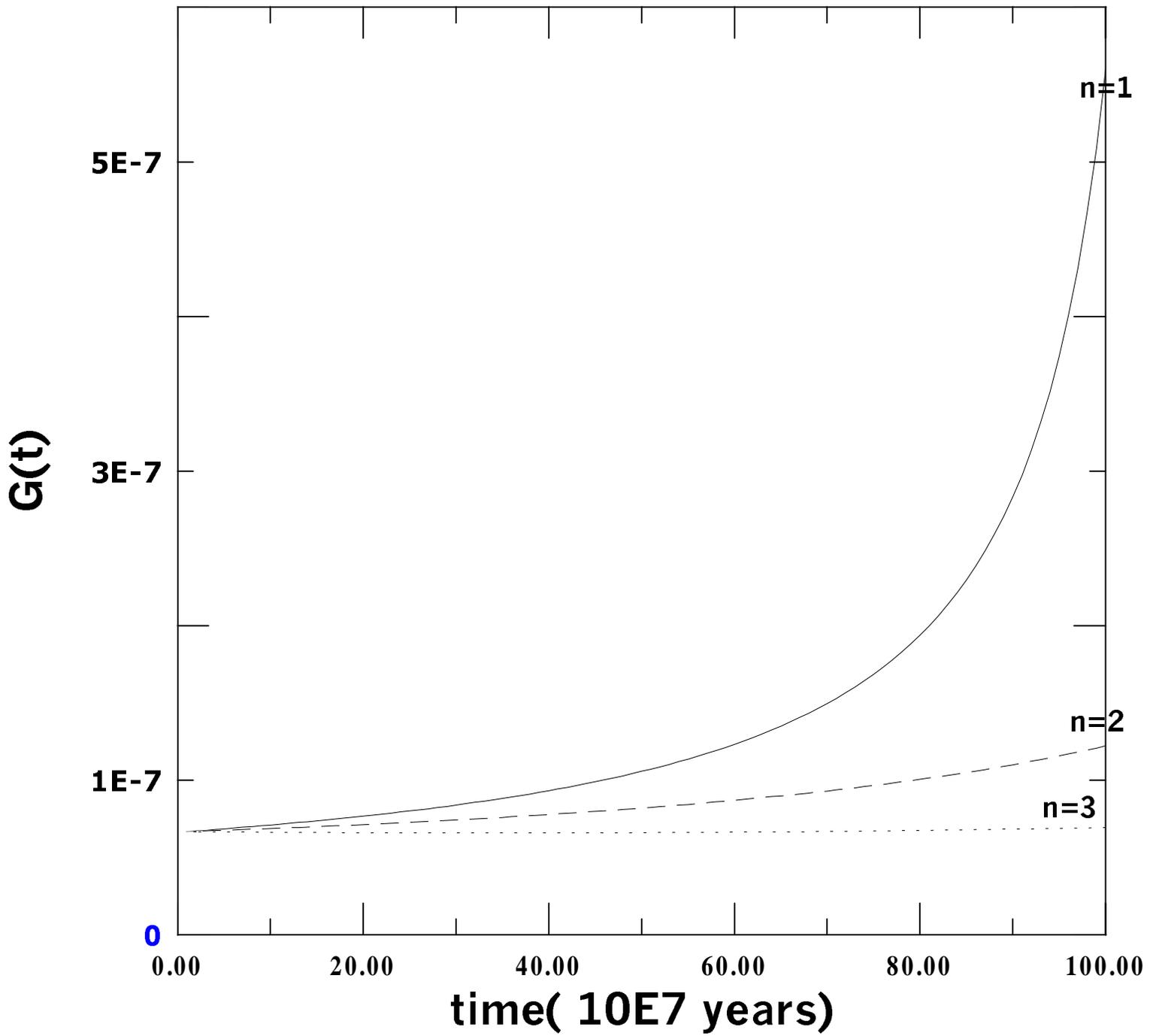
\textbf{Fig 1:G as a time function, when we consider the some ratios between 
}$\phi $\textbf{\ and }$\dot{\phi},$\textbf{\ in this case }$\dot{\phi}/\phi
=10^{-n}$\textbf{, where }$n=1($solid line),$2($long dashed line)\textbf{\
or 3(shor dashed line) . It%
%TCIMACRO{\UNICODE{0xb4}}%
%BeginExpansion
\'{}%
%EndExpansion
s important to observe that in this picture, the time goes from the right to
left, being 0.00 the present time and 100 one billion of years ago. The same
notation is used in the another grafics. }

\begin{center}
\bigskip 
\end{center}

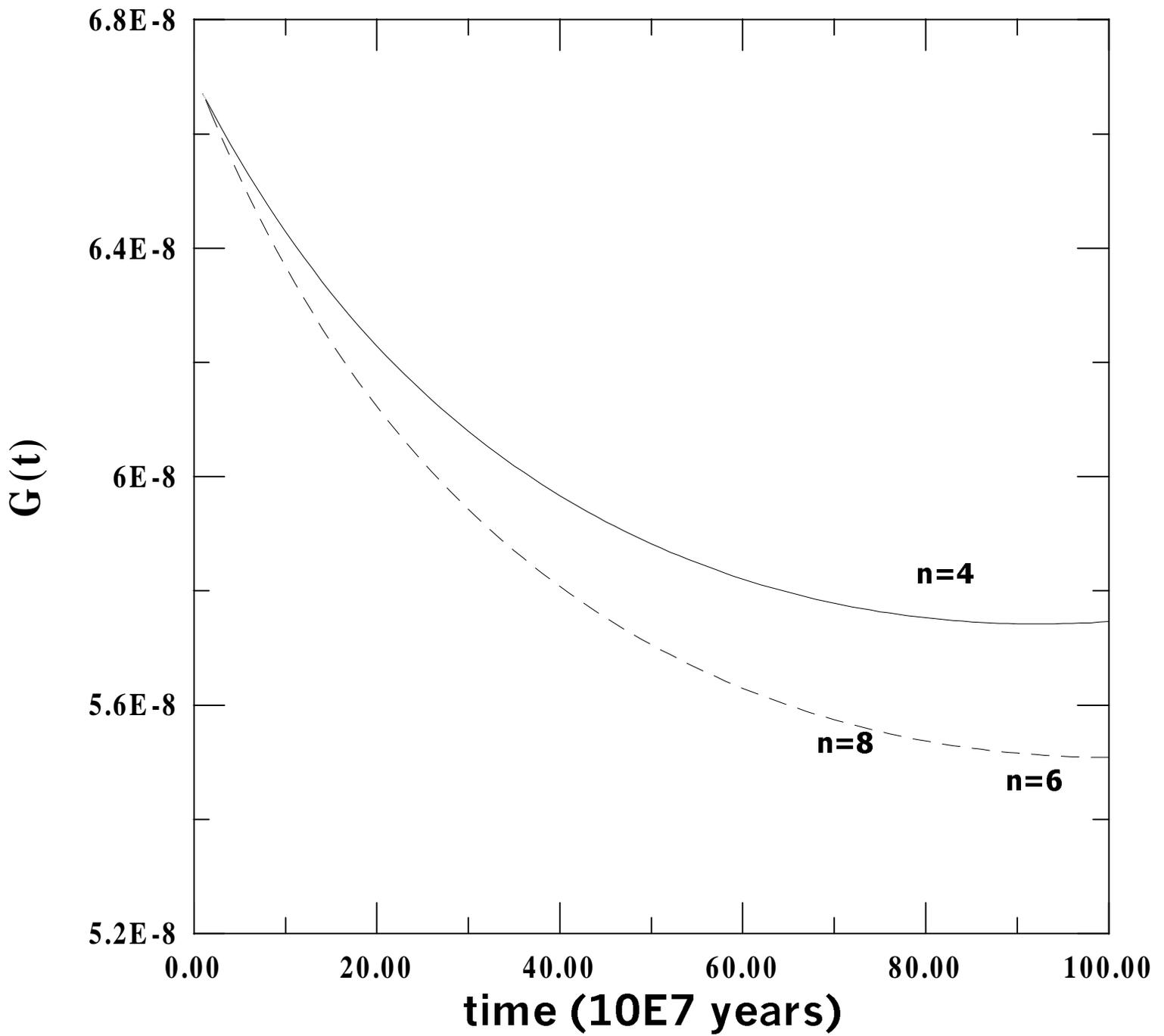
\textbf{Fig 2:\ The case n=4,(solid) 6 (dashed) and 8. }$\dot{\phi}$ \textbf{%
is positive}.\textbf{We can see that n=6 and n=8producesalmost the same
results, so one line superposes the another. We can also see that we have a
growing G(t), and also note this is an acceleratting growing. Due this we
will eliminate these results of a ``fine tunning'' analisys.}

\bigskip 

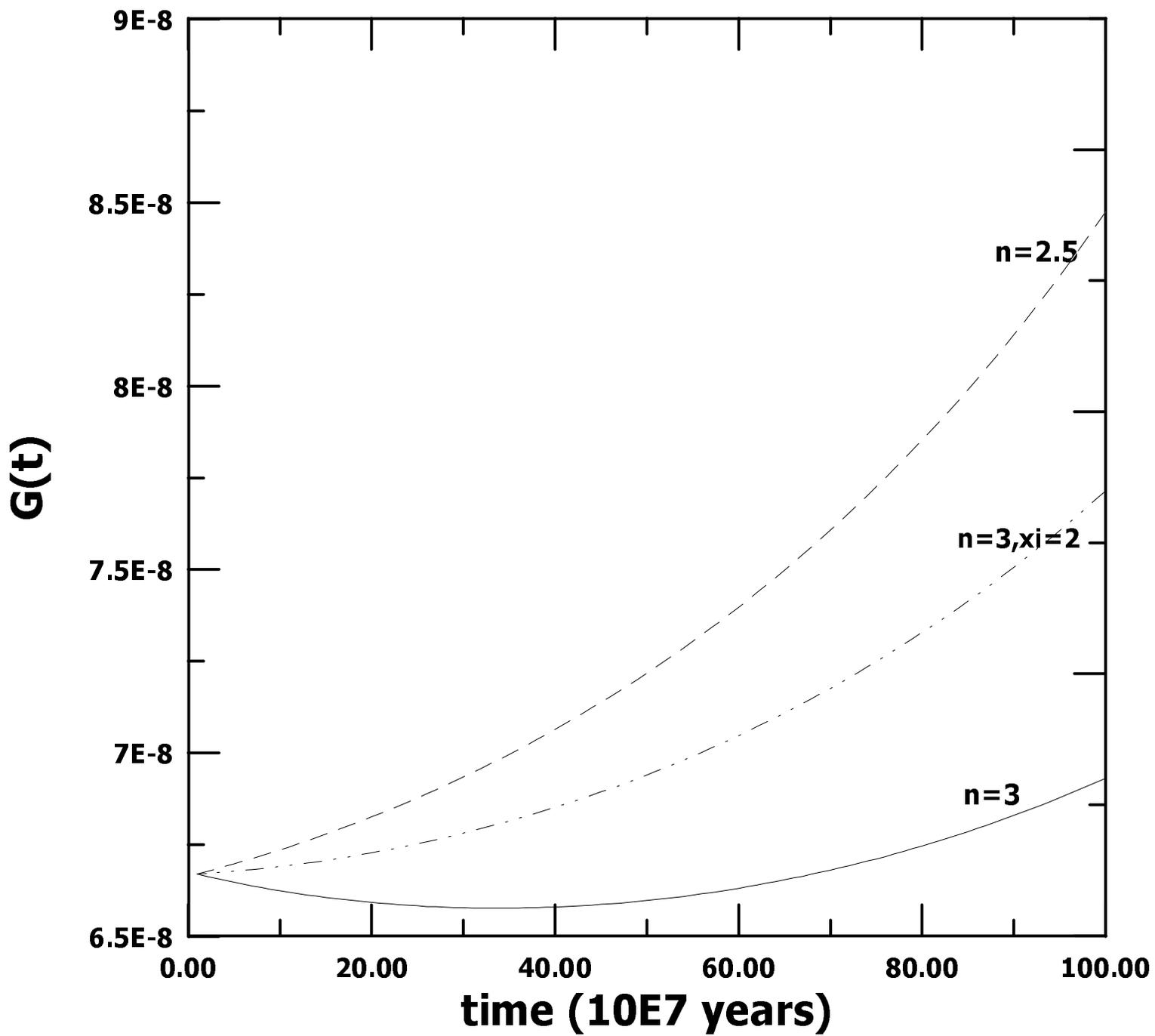
\textbf{Fig 3:A study for the cases with n=2.5(long dashed) and n=3, (}$\dot{%
\phi}=2\times 10^{-3}\phi )$\textbf{(short dashed)}$.$\textbf{For }$\dot{\phi%
}=10^{-3}\phi $\textbf{\ (solid line) we can observe a minimum. We can see
there are an interesting result and worked it as a starting point of a
search for a better fitting of the theorethical parameters.}

\bigskip 

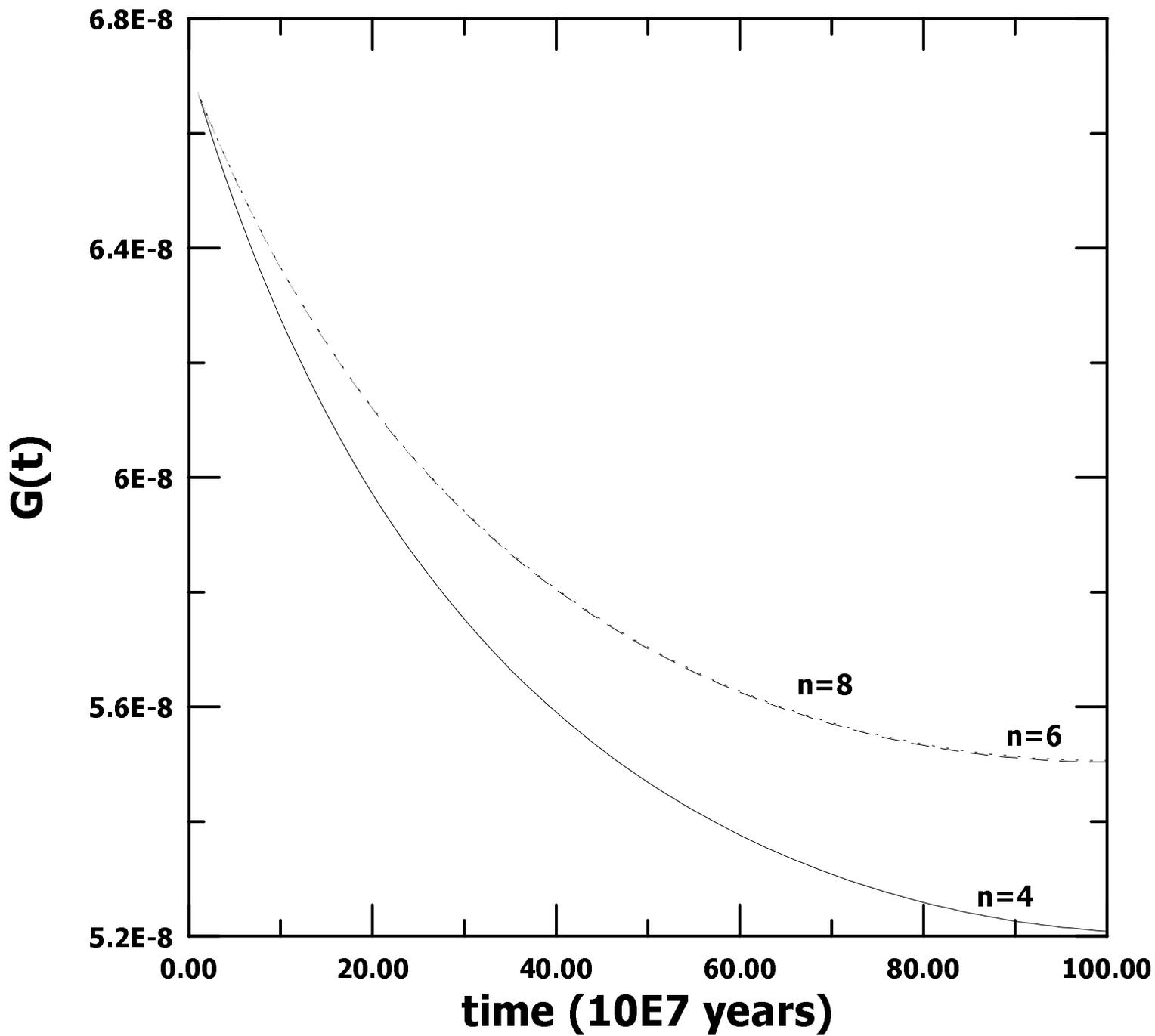
\textbf{Fig 4:G(t) in the cases we have negative values for the intial value
of }$\dot{\phi}.$ \textbf{We use n=4,(solid line) , 6 (long dashed) and 8
(short dashed). Again one result superposes another. The cases n=3 and n=2
have resulted inconsistent values and oscilations.}

\bigskip 

\begin{center}
\bigskip 
\end{center}


\begin{thebibliography}{99}
\bibitem{Dirac}  Dirac,P.A.M.; Nature, \textbf{139, }323 (1937);
Proc.Roy.Soc., \textbf{A165,} 199 (1938).

\bibitem{BD}  Brans, C.; Dicke, R.H., Phys. Rev., \textbf{124,} 925 (1961)

\bibitem{STT}  Barrow, J.D., \textit{Varying G and Other Constants}, at LANL
site

\bibitem{srr}  Hellings, R. Phys. Rev. Lett \textbf{51, }1609 (1983);
Pitjeva, E.V, 
%TCIMACRO{\UNICODE{0xb4}}%
%BeginExpansion
\'{}%
%EndExpansion
%TCIMACRO{\UNICODE{0xb4}}%
%BeginExpansion
\'{}%
%EndExpansion
\textit{Dynamics and Astrometry of Natural and Artificial Celestial
Bodies'', }Kluver Acad. Publ. Nethelands, pg 251 (1997).

\bibitem{llr}  Dickey, J.O, et al., Science \textbf{265, }482 (1994).

\bibitem{BMN}  Bronikov, K.A.;Melnikov, V.N.; Novello, M.; Grav. \& Cosml.
Suppl II, 18 (2002)-gr-qc/0208028.

\bibitem{BP}  Barrow, J.D.; Parsons, P.; Phys. Rev. D, \textbf{55}, 1906
(1997)

\bibitem{Nature}  Bernardis, P.de, et al., Nature, 955, \textbf{404, (}2000).

\bibitem{accelerating}  Riess A.G; et al. Observacional Evidence from
Supernovae for an Accelerating Universe and a Cosmological Constant
-Astronomical Journal astro-ph/9805201.

\bibitem{Will}  Will, C.M., \textit{Theory and Experiment in Gravitational
Physics, }2nd edn, Cambridge UP, Cambridge (1993)

\bibitem{Wein}  Weinberg, S. 
%TCIMACRO{\UNICODE{0xb4}}%
%BeginExpansion
\'{}%
%EndExpansion
\textit{%
%TCIMACRO{\UNICODE{0xb4}}%
%BeginExpansion
\'{}%
%EndExpansion
Gravitation and Cosmology'', Priciples and Aplications of the general Theory
of} \textit{Relativity.},630-631, John Wiley and Sons, New York, (1972)

\bibitem{GWSTT}  Maia, M.R.G.;Barrow, J.D. ; Phys. Rev. D \textbf{50, }6263
(1994).
\end{thebibliography}
\end{document}